\documentclass{ws-ijmpd}
\usepackage{hyperref}
\usepackage[super,compress]{cite}
\begin{document}

\markboth{Hang Wang,etc}
{The equivalence between Einstein and Jordan frames: a study based on the inflationary magnetogenesis model}

\catchline{}{}{}{}{}
%
%%%%%%%%%%%%%%%%%%%%%%%%%%%%%%%%%%%%%%%%%%%%%%%%%%%%%%%%%%%%%%%%%%%%

\title{The equivalence between Einstein and Jordan frames: a study based on the inflationary magnetogenesis model}

\author{Hang Wang\footnote{wang-hang@dlmu.edu.cn},~Shuang Liu\footnote{1120240893shuangliu@dlmu.edu.cn},~Yu Li\footnote{leeyu@dlmu.edu.cn (corresponding author)}~~and~Yao-chuan Wang\footnote{ycwang@dlmu.edu.cn}}

\address{School of Science, Dalian Maritime University, Dalian 116026, China\\
}

%\author{SECOND AUTHOR}
%
%\address{Group, Laboratory, Address\\
%City, State ZIP/Zone, Country\\
%second\_author@group.com}

\maketitle

\begin{history}
\received{Day Month Year}
\revised{Day Month Year}
\end{history}

\begin{abstract}
The equivalence of the Jordan and Einstein frames has been a subject of considerable interest in the field. In this paper, within the context of $f(R)$ gravity, we explore the inflationary magnetogenesis model, focusing on the magnetic field energy density and its spectrum in both the Jordan and Einstein frames to elucidate the equivalence between these two reference frames. Our analysis reveals that during the inflationary epoch, while the magnetic field exhibits a scale-invariant spectrum in the Einstein frame, it demonstrates a blue spectrum in the Jordan frame. Additionally, we investigate the post-inflationary evolution of the magnetic field’s energy density in both frames, uncovering that for scale-invariant spectra in the Einstein frame during inflation, the magnetic field transitions to a blue spectrum, whereas in the Jordan frame, it evolves into a red spectrum. We also establish the conditions under which both frames may exhibit scale-invariant spectra simultaneously during the inflationary period.
\end{abstract}

\keywords{$f(R)$ gravitation; Frame equivalence; Inflationary magnetogenesis model}

\ccode{PACS numbers:}98.80.Cq

%\tableofcontents
\section{Introduction}
In 1998, the first observational evidence of the accelerated expansion of the Universe was obtained\cite{r1,r28}. While Einstein's theory of general relativity provides a fundamental framework for describing gravitational phenomena--successfully explaining the anomalous precession of Mercury's orbit, gravitational lensing effects, and various cosmological observations--it fails to account for critical evolutionary phases of the cosmos, particularly the primordial inflationary epoch and the present-era accelerated expansion.\cite{PhD1}. The introduction of exotic fluids or a cosmological constant represents a conventional approach to bridge this theoretical gap, aligning predictions with observational data\cite{r2}. Nevertheless, these formulation lack fundamental explanations for the underlying mechanisms driving such cosmological phenomena. Consequently, modified gravity theories have been proposed\cite{r3,r4,r25}, including Scalar-Tensor Theories\cite{r5,r6,r7,r29}, $f(R)$ gravity\cite{r8,r9,r10,r24}, Gauss-Bonnet gravity\cite{r11,r12} and $f(
R,T)$ gravity\cite{r13,r14,r15}. Among these, $f(R)$ gravity emerges as the most minimalist extension, expressible through two mathematically equivalent formulations: the Jordan frame and Einstein frame, interconnected via conformal transformations. Although these frames are formally equivalent at the mathematical level, their physical equivalence remains contentious due to differential couplings between the conformal field and matter sectors, phenomenological consequences\cite{m1}.  The establishment of equivalence between the Jordan frame and Einstein frame is essential for preserving the consistency of physical laws\cite{r26}. Should these frames demonstrate mathematical equivalence under conformal transformations, their corresponding formulations should in principle yield consistent descriptions of physical phenomena.

The physical equivalence between the Jordan frame and the Einstein frame can be elucidated from multiple perspectives. For instance, the introduction of a "running unit" has been proposed to demonstrate that they seem to be inequivalent at the quantum level\cite{r16}. Analysis through Noether symmetry further suggest an inability to establish physical equivalence between the two frames in quantum regimes\cite{r18}. Comparative studies of Ricci scalar $R$ reveal distinctions in physical singularity manifestations under different frames, providing another dimension for evaluating their equivalence\cite{m1}. Classical equivalence has been investigated through the Hamiltonian canonical transformations\cite{r31} and J. Velásquez et al. elucidates their equivalence through the generation mechanisms of cosmological primordial magnetic fields.\cite{r7}. However, in the article by J. Velásquez et al., it was established that the coupling functions between electromagnetic and scalar fields in their respective frame adopt power-law forms of their own conformal time. Building upon this foundation, we propose in the present study that the coupling functions bridging these two frames must inherently obey a conformal transformation.

The structure of this paper is organized as follow: In Sec.\ref{s2}, we begin with a brief review of the conformal transformation between the Jordan frame and Einstein frame, followed by a presentation of the action formulations and field equation in both frames. In Sec.\ref{s3}, we devoted to computing the energy density and power spectrum of the primordial magnetic field during the inflationary epoch within both frames. We subsequently discuss the computational results and establish the equivalence conditions between the two formulations. In Sec.\ref{s4}, we extend our investigation to the reheating phase by calculating the corresponding energy density and power spectrum of magnetic fields in both frames, accompanied by comprehensive analysis and discussion of the obtained results. The summary is in Sec.\ref{s5}.

\section{ Field  Equation  in  Jordan  and  Einstein  Frame  \label{s2}}
We consider the action as:
\begin{equation}
\label{e1}
    S=\frac{1}{2\kappa}\int d^4x\sqrt{-g}\,f(R)-\frac{1}{4}\int d^4x\sqrt{-g}\,I(R)F_{\mu\nu}F^{\mu\nu}
\end{equation}
where $\kappa\equiv8\pi G$, $f(R)$ and $I(R)$ are arbitrary functions of $R$, $F_{\mu\nu}$ is EM field tensor. In this paper, we choose the units that $G=c=\hbar=1$ and adopt a metric signature $g_{\mu\nu}=\text{diag}(-,+,+,+)$.

By applying the Legendre transformation to Eq.(\ref{e1}) , we derive the action in Jordan frame:
\begin{eqnarray}
    S_{\rm JF}=\frac{1}{2\kappa}d^4x\sqrt{-g}[\phi R-U(\phi)]-\frac{1}{4}\int d^4x\sqrt{-g}\,I(R(\phi))F^{\mu\nu}F_{\mu\nu}
\end{eqnarray}
where $\phi=\frac{df(R)}{dR}$, $U(\phi)=R(\phi)\phi-f(R(\phi))$. Define $I(R(\phi))$ as a function of $\phi$, i.e.$I(R(\phi))=I(\phi)$. We assume that the scalar field $\phi$ is the field responsible for inflation.

Varying the action $S_{\rm JF}$ with respect to the metric tensor  $g^{\mu\nu}$, we obtain the field equation of gravitation:
\begin{eqnarray}
    G_{\mu\nu}=\kappa T_{\mu\nu}^{(\phi)}+\frac{\kappa}{\phi}T_{\mu\nu}^{\rm EM}
\end{eqnarray}
where
\begin{eqnarray}
     T_{\mu\nu}^{(\phi)}=\frac{1}{\kappa\phi}(\nabla_{\mu}\nabla_{\nu}\phi-g_{\mu\nu}\square\phi)-\frac{1}{2\kappa\phi}g_{\mu\nu}U(\phi)
\end{eqnarray}
\begin{eqnarray}
    T_{\mu\nu}^{\rm EM}=I^2(\phi)\left[g^{\alpha\beta}F_{\mu\alpha}F_{\nu\beta}-g_{\mu\nu}\frac{F_{\alpha\beta}F^{\alpha\beta}}{4}\right]
\end{eqnarray}
here $\square=\nabla_{\mu}\nabla^{\mu}$.

Varying the action $S_{\rm JF}$ with respect to the scalar field $\phi$ and the EM 4-potential $A_{\mu}$, respectively. We get

Klein-Gordon equation:
\begin{eqnarray}
    R=\frac{dU(\phi)}{d\phi}+\kappa I(\phi)\frac{dI(\phi)}{d\phi}F_{\mu\nu}F^{\mu\nu}
\end{eqnarray}

And Maxwell's equation:
\begin{equation}
    \label{e7}
    \frac{1}{\sqrt{-g}}\partial_{\mu}[\sqrt{-g}I^2(\phi)F^{\mu\nu}]=0
\end{equation}

In this context, we perform a conformal transformation of the metric to derive the action in the Einstein frame. The conformal transformation is
\begin{eqnarray}
    \tilde{g}_{\mu\nu}=f_{R}~g_{\mu\nu}=\phi g_{\mu\nu},~~~~\tilde{\phi}\equiv\sqrt{\frac{3}{2\kappa}}\ln\phi
\end{eqnarray}
where $\tilde{\phi}$ is conformal field, $f_{R}$ is the first derivative of f(R) with respect to R.

The Ricci scalars R and $\tilde{R}$ which is associated with $\tilde{g}_{\mu\nu}$ have the following relationship
\begin{equation}
    \label{e9}
    R=\phi\tilde{R}+6\tilde{g}^{\mu\nu}\sqrt{\phi}(\tilde{\nabla}_{\mu}\tilde{\nabla}_{\nu}\sqrt{\phi})-12\tilde{g}^{\mu\nu}(\tilde{\nabla}_{\mu}\sqrt{\phi})(\tilde{\nabla}_{\nu}\sqrt{\phi})
\end{equation}

Utilizing Eq.(\ref{e9}) and $\sqrt{-\tilde{g}}=\phi^2\sqrt{-g}$, the action within the Einstein frame can be calculated
\begin{eqnarray}
    S_{\rm EF}=\int\sqrt{-\tilde{g}}\left[\frac{\tilde{R}}{2\kappa}-\frac{1}{2}\tilde{g}^{\mu\nu}(\tilde{\nabla}_{\mu}\tilde{\phi})(\tilde{\nabla}_{\nu}\tilde{\phi})-V(\tilde{\phi})\right]\\-\frac{1}{4}\int \sqrt{-\tilde{g}}~d^4xK^2(\tilde{\phi})\tilde{F}_{\mu\nu}\tilde{F}^{\mu\nu}
\end{eqnarray}
where $V(\tilde{\phi})=\frac{U(\phi)}{2\kappa\phi^2}$, $K(\tilde{\phi})$ is the function obtained after applying a conformal transformation to $I(\phi)$.

Then, according to $\tilde{\nabla}_{\alpha}V_{\beta}=\nabla_{\alpha}V_{\beta}+(g_{\alpha\beta}g^{\mu\nu}\nabla_{\nu}\ln\sqrt{\phi}-2\delta^{\mu}_{(\alpha}\nabla_{\beta)}\ln\sqrt{\phi})V_{\mu}$, we find that the EM field tensor $F_{\mu\nu}$ is invariant under conformal transformations. Similarly, we obtain the evolution equations of the fields within the Einstein frame.

Equation of the gravitational field:
\begin{eqnarray}
     \tilde{G}_{\mu\nu}=\kappa(\tilde{T}_{\mu\nu}^{(\tilde{\phi})}+\tilde{T}_{\mu\nu}^{\rm EM})
\end{eqnarray}
\begin{eqnarray}
    \tilde{T}_{\mu\nu}^{(\tilde{\phi})}=\tilde{\nabla}_{\mu}\tilde{\phi}\tilde{\nabla}_{\nu}\tilde{\phi}-\frac{1}{2}\tilde{g}_{\mu\nu}\tilde{g}^{\alpha\beta}\tilde{\nabla}_{\alpha}\tilde{\phi}\tilde{\nabla}_{\beta}\tilde{\phi}-\tilde{g}_{\mu\nu}V(\tilde{\phi})
\end{eqnarray}
\begin{eqnarray}
     \label{e14}
     \tilde{T}_{\mu\nu}^{\rm EM}=K^2\left[\tilde{g}^{\alpha\beta}F_{\mu\alpha}F_{\nu\beta}-\tilde{g}_{\mu\nu}\frac{F_{\alpha\beta}F^{\alpha\beta}}{4}\right]
\end{eqnarray}

Equation of scalar field $\tilde{\phi}$:
\begin{eqnarray}
    \tilde{\square}\tilde{\phi}-\frac{dV(\tilde{\phi})}{d\tilde{\phi}}=\frac{K}{2}\frac{dK}{d\tilde{\phi}}F_{\mu\nu}F^{\mu\nu}
\end{eqnarray}

And Maxwell's equation:
\begin{equation}
    \label{e16}
    \frac{1}{\sqrt{-\tilde{g}}}\tilde{\partial}_{\mu}[\sqrt{-\tilde{g}}K^2(\tilde{\phi})F^{\mu\nu}]=0
\end{equation}

Then, we consider the spatially flat FRW metric as background in Jordan frame
\begin{equation}
    \begin{split}
        ds^2&=-dt^2+a^2(t)[dx^2+dy^2+dz^2]\\&=a^2(\eta)[-d\eta^2+dx^2+dy^2+dz^2]
    \end{split}
\end{equation}
where t is cosmic time, $\eta=\int\frac{dt}{a}$ is the conformal time.

Thus, in the Einstein frame, we have
\begin{equation}
    \begin{split}
        d\tilde{s}^2&=\phi\{-dt^2+a^2(t)[dx^2+dy^2+dz^2]\}\\&=-d\tilde{t}^2+\tilde{a}^2(\tilde{t})[dx^2+dy^2+dz^2]\\&=\tilde{a}^2(\tilde{\eta})[-d\tilde{\eta}^2+dx^2+dy^2+dz^2]
    \end{split}
\end{equation}
Here, $\tilde{t}$ is cosmic in Einstein frame and $\tilde{\eta}=\int\frac{d\tilde{t}}{\tilde{a}}$ is the conformal time in Einstein frame.

\section{Magnetic Field Generation During Inflation \label{s3}}

Now, we will separately discuss the magnetogenesis model in the two respective frames. The action in the Einstein frame consists of a standard kinetic term plus a potential term, which facilitates the discussion. Therefore, we first investigate the problem within the Einstein frame. In this paper, we consider a simple $f(R)$ model-Starobinsky model in the Jordan frame.

\subsection{Magnetogenesis model in Einstein frame}

We adopt the Coulomb gauge as follow
\begin{eqnarray}
    A_{0}(\tilde{\eta},\textbf{x})=0,~~~~\tilde{\partial}_{j}A^{j}(\tilde{\eta},\textbf{x})=0
\end{eqnarray}

Then, Eq.(\ref{e16}) can be expressed as,
\begin{eqnarray}
    \label{e20}
    \tilde{\partial}_{0}\tilde{\partial}^{0}A_{i}+2\frac{\tilde{\partial}_{0}K}{K}\tilde{\partial}_{0}A_{i}-\tilde{a}^2\tilde{\partial}_{j}\tilde{\partial}^{j}A_{i}=0
\end{eqnarray}
where $\tilde{\partial}^{j}$ is defined as $\tilde{\partial}^{j}=\tilde{g}^{jk}\tilde{\partial}_{k}=\tilde{a}^{-2}\delta^{jk}\tilde{\partial}_{k}$.

We promote $A_{i}$ to an operator and impose the quantization condition, then expand $A_{i}$ in terms of creation and annihilation operators $\hat{b}^{\dagger}(k)$ and $\hat{b}(k)$ with the co-moving wave vector in Fourier space,
\begin{eqnarray}
    \label{e21}
     A_{i}(\textbf{x},\tilde{\eta})=\int\frac{d^3k}{(2\pi)^3}\sum_{\lambda=1}^{2} e_{\lambda,i}(\textbf{k})[\hat{b}_{\lambda}(\textbf{k})A(k,\tilde{\eta})\hat{e}^{i\textbf{k}\cdot\textbf{x}}+\hat{b}^{\dagger}_{\lambda}(\textbf{k})A^{\ast}(k,\tilde{\eta})e^{-i\textbf{k}\cdot\textbf{x}}]
\end{eqnarray}
where $e_{\lambda,i}$ are the polarization vectors.

We substitute the Eq.(\ref{e21}) into Eq.(\ref{e20}) and define a new variable $\bar{A}\equiv\tilde{a}A(k,\tilde{\eta})$, to get 
\begin{eqnarray}
    \tilde{\partial}_{0}\tilde{\partial}^{0}\bar{A}_{i}+2\frac{\tilde{\partial}_{0}K}{K}+k^2\bar{A}_{i}=0
\end{eqnarray}

One can also define $\mathcal{A}=K(\tilde{\eta})\bar{A}$ in order to eliminate the first derivative term, to get
\begin{eqnarray}
    \label{e23}
     \tilde{\partial}_{0}\tilde{\partial}^{0}{\mathcal{A}}(k,\tilde{\eta})+\left(k^2-\frac{{\tilde{\partial}_{0}\tilde{\partial}^{0}K}}{K}\right)\mathcal{A}(k,\tilde{\eta})=0
\end{eqnarray}

We assume that the coupling function $K$ evolves as $\tilde{a}$ power law during the inflation,
\begin{eqnarray}
    K(\tilde{a})=K_{i}\left(\frac{\tilde{a}}{\tilde{a}_{i}}\right)^{-\alpha}
\end{eqnarray}
where $K_{i}$ is a constant, $\tilde{a}_{i}$ is the scalar factor at the beginning of inflation in Einstein frame.

And we have the evolution of the scale factor as\cite{r8}
\begin{eqnarray}
    \tilde{a}(\tilde{t})\simeq\left(1-\frac{M^2}{12H_{i}^2}M\tilde{t}\right)~\tilde{a}_{i}~e^{M\tilde{t}/2}
\end{eqnarray}
where $H_{i}$ is the Hubble parameter at the beginning of inflation in Jordan frame. We assume $t_{i}=0$ and  $a(t_{i})=1$.Since the Hubble parameter $H$ was extremely large at the onset of inflation, the scalar factor $\tilde{a}$ here can be approximated as undergoing quasi-exponential inflation, we have $\tilde{a}\tilde{H}\rightarrow-1/\tilde{\eta}$. Thus we obtain\cite{r17}
\begin{eqnarray}
    \label{e26}
    K(\tilde{a})=K(\tilde{\eta})\propto\tilde{\eta}^{\alpha}
\end{eqnarray}

Then Eq.(\ref{e23}) reduces to,
\begin{eqnarray}
    \tilde{\partial}_{0}\tilde{\partial}^{0}{\mathcal{A}}(k,\tilde{\eta})+\left[k^2-\frac{\alpha(\alpha-1)}{\tilde{\eta}^2}\right]\mathcal{A}(k,\tilde{\eta})=0
\end{eqnarray}

The solution of this equation can be written as
\begin{eqnarray}
    \label{e28}
    \mathcal{A}_{1}(k,\tilde{\eta})=\sqrt{-k\tilde{\eta}}~\left[c_{1}(k)J_{\alpha-\frac{1}{2}}(-k\tilde{\eta})+c_{2}(k)J_{\alpha+\frac{1}{2}}(k\tilde{\eta})\right]
\end{eqnarray}
where $J$ is Bessel function of the first kind, constants $c_{1}(k)$ and $c_{2}(k)$ are fixed by B-D vacuum condition in the limit of $(-k\tilde{\eta})\rightarrow\infty$,
\begin{eqnarray}
    c_{1}(k)=\sqrt{\frac{\pi}{4k}}\frac{\exp(-i\pi\alpha/2)}{\cos(\pi\alpha)},~~~~c_{2}(k)=\sqrt{\frac{\pi}{4k}}\frac{\exp[i\pi(\alpha+1)/2]}{\cos(\pi\alpha)}
\end{eqnarray}

Using Eq.(\ref{e14}), we get spectral energy density in the magnetic field\cite{r19},
\begin{eqnarray}
    \label{e30}
    \frac{d\rho_{B}}{d\ln k}=\frac{1}{2\pi^2}\frac{k^5}{\tilde{a}^4}\left|\mathcal{A}(k,\tilde{\eta})\right|^2
\end{eqnarray}

We substitute Eq.(\ref{e28}) into the above equation,in the super-horizon limit $(-k\tilde{\eta})\ll1$,we get
\begin{eqnarray}
    \frac{d\rho_{B}}{d\ln k}\approx\frac{\mathcal{C}^2(n)}{2\pi^2}\tilde{H}_{f}^4(-k\tilde{\eta})^{2n+4}
\end{eqnarray}
where $\tilde{H}_{f}$ is the Hubble parameter during inflation in Einstein frame, $n=\alpha$ if $\alpha<1/2$ and $n=1-\alpha$ for $\alpha\ge1/2$, and
\begin{eqnarray}
    \mathcal{C}(n)=\frac{\pi}{2^{n-1/2}}\frac{e^{-i\pi n/2}}{\Gamma(n+1/2)\cos(\pi n)}.
\end{eqnarray}
where $\Gamma$ is the Gamma function. 

Thus we find that scale invariance magnetic field spectrum $(2n+4=0)$ have two possible values of $\alpha$,
\begin{eqnarray}
    \alpha=3~~~,~~~ \alpha=-2.
\end{eqnarray}

And, we define the number of e-foldings from $\tilde{t}=\tilde{t}_{i}$ to $\tilde{t}$ in the Einstein frame, then we get\cite{r8}\cite{PhD1}
\begin{eqnarray}
    \label{e34}
    \ln\left(\frac{\tilde{a}}{\tilde{a}_{i}}\right)\equiv\tilde{N}\equiv\int_{\tilde{t}_{i}}^{\tilde{t}}\tilde{H} ~d\tilde{t}\simeq\frac{3}{4}e^{\sqrt{2\kappa/3}\tilde{\phi}}
\end{eqnarray}
where $\tilde{t}_{i}$ represent the cosmic time corresponding to the beginning of inflation.

Subsequently, we formulate the magnetogenesis model in the Jordan frame to contrast the two frames, with the reheating epoch analysis reserved for Section \ref{s4}.

\subsection{Magnetogenesis model in Jordan frame}

By following the computational procedures outlined in the previous section, we derive the dynamical equation governing the electromagnetic field,
\begin{eqnarray}
    \partial_{0}\partial^{0}A_{i}+2\frac{\partial_{0}I}{I}\partial_{0}A_{i}-a^2\partial_{j}\partial^{j}A_{i}=0
\end{eqnarray}
where $\partial^{j}$ is defined as $\partial^{j}=g^{jk}\partial_{k}=a^{-2}\delta^{jk}\partial_{k}$.

Similarly, the evolution of the mode function $\mathcal{A}$ is also given by
\begin{eqnarray}
    \label{e36}
    \partial_{0}\partial^{0}\mathcal{A}(k,\eta)+\left(k^2-\frac{\partial_{0}\partial^{0}I}{I}\right)\mathcal{A}=0
\end{eqnarray}
where $\mathcal{A}=a(\eta)I(\eta)A(k,\eta)$. We also get for the spectral energy densities in the magnetic fields,
\begin{eqnarray}
     \label{e37}
     \frac{d\rho_{B}}{d\ln k}=\frac{1}{2\pi^2}\frac{k^5}{a^4}\left|\mathcal{A}(k,\eta)\right|^2
\end{eqnarray}

We find that the equation of motion for the electromagnetic field and magnetic spectrum exhibit identical forms within both frames, with the distinction residing exclusively in the coupling functions. In the following, we will give the form of the coupling function in the Jordan frame through the form of the coupling function in the Einstein frame. However, the direct application of $\phi(\eta)$ and $\tilde{\phi}(\tilde{\eta})$ to transform $\tilde{\eta}$ into $\eta$ results in excessive complexity; therefore, we introduce the number of e-foldings as a simplifying mechanism.

According to Eq.(\ref{e34}) and the function between $\phi$ and $\tilde{\phi}$\cite{r8}, we get 
\begin{eqnarray}
    \tilde{N}=ln\left(-\frac{1}{\tilde{H}\tilde{\eta}}\right)=\frac{3}{4}\phi
\end{eqnarray}
Thus, we obtain
\begin{eqnarray}
    \label{e39}
    K(\tilde{\eta})\propto\tilde{\eta}^{\alpha}=(-\tilde{H})^{\alpha}(e^{3\phi/4})^{\alpha}
\end{eqnarray}

We also define the number of e-foldings in the Jordan frame from $t_{i}=0$ to $t$\cite{r8}:
\begin{eqnarray}
    \label{e40}
    ln\left(\frac{a(\eta)}{a(\eta_{i})}\right)\equiv N\equiv\int_{t_{i}=0}^{t}H~dt\simeq H_{i}t-\frac{M^2}{12}t^2,
\end{eqnarray}
and according to the evolutionary equation for $\phi$ \cite{r8}
\begin{eqnarray}
    \phi(t)\simeq\frac{4}{M^2}\left(H_{i}-\frac{M^2}{6}t\right)^2,
\end{eqnarray}
We get the relationship between $\phi$ and $N$ as,
\begin{eqnarray}
    \label{e42}
    \phi(N)=\frac{4H_{i}^2}{M^2}-\frac{4}{3}N
\end{eqnarray}

By substituting Eq.(\ref{e42}) into Eq.(\ref{e39}), we obtain
\begin{eqnarray}
    K(\tilde{\eta})=(-\tilde{H})^{\alpha}~e^{-N\alpha}~e^{\frac{3H_{i}}{M^2}\alpha}
\end{eqnarray}
which combine the Hubble parameter in the Einstein frame\cite{r8}
\begin{eqnarray}
    \tilde{H}(\tilde{t})\simeq\frac{M}{2}\left[1-\frac{M^2}{6H_{i}^2}\left(1-\frac{M^2}{12H_{i}^2}M\tilde{t}\right)^{-2}\right],
\end{eqnarray}
we get
\begin{eqnarray}
    \label{e45}
    K(\tilde{\eta})=D\cdot[g(N)h(N)]=I(\eta)=I(N)
\end{eqnarray}
where $D=(-\frac{M}{2}e^{3H_{i}^2/M^2})^{\alpha}$ is a coupling constant, $g(N)=[1-\frac{M^2}{6H_{i}^2}(1-\frac{M^2}{6H_{i}^2})^{-2}]^{\alpha}$, $h(N)=e^{-N\alpha}$.

In the $R^2$ inflation model, the scalar factor in Jordan frame written as,\cite{r8}
\begin{eqnarray}
    a(t)\simeq a_{i}e^{H_{i}t-\frac{M^2}{12}t^2}
\end{eqnarray}
which is also regarded as quasi-exponential inflation. That means $a(\eta)\rightarrow-1/H\eta$. We assume $a(\eta_{i})=1$ and substituting $a(\eta)$ into Eq.(\ref{e40}) and differentiating both sides twice, yields that, $\partial_{\eta\eta}I(\eta)=\eta^{-2}\partial_{NN}I(N)$. Thus, the function $Y(N)$ can be written as,
\begin{eqnarray}
    \label{e47}
    Y(N)=\frac{\partial_{\eta\eta}I}{I}=\frac{\partial_{NN}I(N)}{\eta^2I(N)}=\frac{1}{\eta^2}\left\{\alpha^2-2\alpha\frac{\partial_{N}[g(N)]}{g(N)}+\frac{\partial_{NN}[g(N)]}{g(N)}\right\}
\end{eqnarray}
where \begin{eqnarray}
    \label{e48}
    \frac{\partial_{N}[g(N)]}{g(N)}=\frac{-\alpha\left[\frac{M^4}{18H_{i}^4}(1-\frac{M^2}{6H_{i}^2}N)^{-3}\right]}{1-\frac{M^2}{6H_{i}^2}(1-\frac{M^2}{6H_{i}^2}N)^{-2}}\equiv a
\end{eqnarray}
\begin{eqnarray}
    \label{e49}
    \frac{\partial_{NN}[g(N)]}{g(N)}=\frac{\alpha(\alpha-1)\left[\frac{M^4}{18H_{i}^4}(1-\frac{M^2}{6H_{i}^2}N)^{-3}\right]^2}{\left[1-\frac{M^2}{6H_{i}^2}(1-\frac{M^2}{6H_{i}^2}N)^{-2}\right]^2}-\frac{\alpha\frac{M^6}{36H_{i}^6}(1-\frac{M^2}{6H_{i}^2}N)^{-4}}{1-\frac{M^2}{6H_{i}^2}(1-\frac{M^2}{6H_{i}^2}N)^{-2}}\equiv b
\end{eqnarray}

Moreover, due to the logarithmic relationship between $N$ and $\eta$, where N is sufficiently small relative to $\eta$, it can be considered as a quasi-constant\cite{PhD1}. Then, plugging Eqs.(\ref{e47}-\ref{e49}) into Eq.(\ref{e36}), the solution to this equation is given by
\begin{eqnarray}
    \label{e50}
    \mathcal{A}_{1}(k,\eta)=\sqrt{-k\eta}[C_{1}(k)J_{\nu}(-k\eta)+C_{2}(k)J_{-\nu}(-k\eta)]
\end{eqnarray}
where $\nu=\frac{1}{2}\sqrt{4\alpha^2-8a\alpha+4b+1}$, 
and 
\begin{equation}
     \begin{cases}
            C_1(k)=\sqrt{\frac{\pi}{4k}}\frac{e^{-i\pi(\nu+\frac{1}{2})/2}}{\cos[\pi(\nu+1/2)]}
            \\
            C_2(k)=\sqrt{\frac{\pi}{4k}}\frac{e^{i\pi(\nu+\frac{3}{2})/2}}{\cos[\pi(\nu+1/2)]}
        \end{cases}
\end{equation}

In the super-horizon limit $-k\eta\ll1$, the Eq.(\ref{e50}) becomes\cite{r19},
\begin{eqnarray}
    \label{e52}
    \lim_{-k\eta\rightarrow0}\mathcal{A}_{1}(k,\eta)=\sqrt{\frac{1}{k}}\left[D_{1}(\nu)(-k\eta)^{\nu+\frac{1}{2}}+D_{2}(\nu)(-k\eta)^{\frac{1}{2}-\nu}\right]
\end{eqnarray}
where 
\begin{equation}
    \begin{cases}
        D_{1}(\nu)=\frac{\sqrt{\pi}}{2^{\nu+1}}\frac{e^{-i\pi(\nu+\frac{1}{2})/2}}{\Gamma(\nu+1)\cos[\pi(\nu+1/2)]}
        \\
        D_{2}(\nu)=\frac{\sqrt{\pi}}{2^{-\nu}}\frac{e^{i\pi(\nu+\frac{3}{2})/2}}{\Gamma(-\nu+1)\cos[\pi(\nu+1/2)]}
    \end{cases}
\end{equation}

By substituting Eq.(\ref{e52}) into Eq.(\ref{e37}),  we also get the spectral energy density in the magnetic field,
\begin{eqnarray}
    \frac{d\rho_{B}}{d\ln k}=\frac{\mathcal{D}(m)}{2\pi^2}H_{f}^4(-k\eta)^{2m+4}
\end{eqnarray}
where $H_{f}$ is the Hubble parameter during inflation in the Jordan frame, $m=\nu+\frac{1}{2}$ if $\nu<0$ and $m=\frac{1}{2}-\nu$ for $\nu\ge0$.

We also find that scale invariance magnetic field spectrum:
\begin{eqnarray}
    \nu=\pm\frac{5}{2},
\end{eqnarray}
thus we obtain,
\begin{eqnarray}
    \alpha\approx\pm2.44949
\end{eqnarray}
here, we used the value that $H_i\simeq 10^{16}\mathrm{GeV}$.

Substituting the value of $\alpha$ in the Einstein frame into equation $\nu=\frac{1}{2}\sqrt{\alpha^2-8a\alpha+4b+1}$ yields,
\begin{equation}
    \begin{split}
        &(1).~~\nu\approx2.06155\Rightarrow~2m+4\approx0.8769.~~~~~~(\alpha=-2)\\&(2).~~\nu\approx3.04138\Rightarrow~2m+4\approx-1.08276.~~(\alpha=3)
    \end{split}
\end{equation}

This implies that a scalar invariant magnetic energy spectrum in the Jordan frame would undergo redshift($\alpha=3$) or blueshift($\alpha=-2$) when transformed into the Einstein frame.

Given the requirement that the magnetic fields in both frames exhibit scalar invariance spectrum, we must satisfy the following relation from the coupling function between two frames,
\begin{eqnarray}
    \alpha^2-2a\alpha+b=\alpha^2+\alpha\Rightarrow\alpha=-\frac{b}{2a+1}.
\end{eqnarray}
Since the values of $a$ and $b$ are negligible, $\alpha$ can be considered approximately 0. i.e.~$\nu\simeq0$. That is to say, it is impossible for the scalar invariance spectrum of the magnetic field to appear simultaneously in the two frames, unless the coupling function is a constant. However, in this way, the conformal invariance of the electromagnetic field will not be broken, and thus the primordial magnetic field cannot be generated through inflation.

\section{Inflationary magnetogenesis in the post inflationary epochs\label{s4}}

In the previous section, we presented the energy density and power spectrum of magnetic fields in both the Jordan frame and Einstein frame during inflation. However, the spectral index corresponding to the scalar-invariant spectrum in the Einstein frame, characterized by $\alpha=-2$ , leads to a backreaction problem\cite{r22,r23}. Therefore, in this section, we focus on the case of $\alpha=-2$ and separately discuss the post-inflationary evolution of magnetic fields within these two frames.

\subsection{Magnetic energy density in the Einstein frame after inflation}

We consider the coupling function as\cite{r20},
\begin{equation}
K=\begin{cases}
    K_{1}&=\left(\frac{\tilde{a}}{\tilde{a}_{i}}\right)^{-\alpha},
\qquad
\tilde{a}_{i}<\tilde{a}<\tilde{a}_{f}~.
\\
K_{2}&=\left(\frac{\tilde{a}_{f}}{\tilde{a}_{i}}\right)^{-\alpha}\left(\frac{\tilde{a}}{\tilde{a}_{f}}\right)^{-\beta},
 \qquad
 \tilde{a}_{f}<\tilde{a}<\tilde{a}_{r}~.
\end{cases}
\end{equation}
here $\tilde{a}_{r}$ is the scale factor in the Einstein frame at the end of reheating.

Since calculating the power spectrum of the magnetic field only requires  the solution in the super-horizon scales, substituting $K_{2}$ into Eq.(\ref{e23}) yields a solution of the form\cite{r20,r19}:
\begin{eqnarray}
    \bar{A}_{2}(k,\tilde{\eta})=\frac{\mathcal{A}_{2}(k,\tilde{\eta})}{K_{2}}=d_{1}+d_{2}\int_{\tilde{\eta}_{f}}^{\tilde{\eta}}\frac{1}{K^2_{2}}d\tilde{\eta}.
\end{eqnarray}
where $\tilde{\eta}_{f}$ is the conformal time at the end of inflation in the Einstein frame. 

Then, we impose the continuity of $\tilde{a}$ and $\tilde{\partial}_{0}\tilde{a}$ at $\tilde{\eta}=\tilde{\eta}_{f}$, yielding the expression for $\tilde{a}$ as,
\begin{eqnarray}
    \tilde{a}(\tilde{\eta})=\frac{\tilde{H}_{f}\tilde{a}_{f}^3}{4}\left(\tilde{\eta}+\frac{3}{\tilde{a}_{f}\tilde{H}_{f}}\right)^2
\end{eqnarray}

Thus, we get the solution denoted by $\tilde{a}$,
\begin{eqnarray}
    \bar{A}_{2}(k,\tilde{\eta})=d_{1}+d_{2}\int_{\tilde{a}_{f}}^{\tilde{a}}\frac{1}{\tilde{a}_{f}^2\tilde{H}_{f}\sqrt{\tilde{a}/\tilde{a}_{f}}K_{2}^2}d\tilde{a}
\end{eqnarray}

We also obtain the expression of $\bar{A}_{1}(k,\tilde{\eta})$ by expanding  Eq.(\ref{e28}) in the super-horizon limit,
\begin{equation}
   \begin{split}
       \bar{A}_{1}(k,\tilde{\eta})=&c_{1}\frac{2^{\frac{1}{2}-\alpha}k^{\alpha}\tilde{H}_{f}^{-\alpha}}{\Gamma(\frac{1}{2}+\alpha)}\left[1-\frac{\left(\frac{k}{\tilde{a}\tilde{H}_{f}}\right)^2}{4(\frac{1}{2}+\alpha)}-\frac{\left(\frac{k}{\tilde{a}\tilde{H}_{f}}\right)^4}{32(\frac{1}{2}+\alpha)(\frac{3}{2}+\alpha)}\right]\\&+c_{2}\frac{k^{\alpha}\tilde{H}_{f}^{-\alpha}\left(\frac{k}{\tilde{a}\tilde{H}_{f}}\right)^{-2\alpha+1}}{2^{-\alpha+\frac{1}{2}}\Gamma(\frac{3}{2}-\alpha)}
   \end{split}
\end{equation}
here we retained the higher order terms of $J_{\alpha-1/2}$. In order to fix the constants $d_{1}$ and $d_{2}$, we demand that $\bar{A}_{1}(k,\tilde{\eta})$, $\bar{A}_{2}(k,\tilde{\eta})$ and their first derivatives are equal at $\tilde{\eta}=\tilde{\eta}_{f}$.Thus, we get,
\begin{equation}
        \begin{cases}
            d_{1}=c_{1}\frac{2^{\frac{1}{2}-\alpha}k^{\alpha}\tilde{H}_{f}^{-\alpha}}{\Gamma(\frac{1}{2}+\alpha)}\left[1-\frac{\left(\frac{k}{\tilde{a}_{f}\tilde{H}_{f}}\right)^2}{4(\frac{1}{2}+\alpha)}-\frac{\left(\frac{k}{\tilde{a}_{f}\tilde{H}_{f}}\right)^4}{32(\frac{1}{2}+\alpha)(\frac{3}{2}+\alpha)}\right]
            \\
            \qquad+c_{2}\frac{k^{\alpha}\tilde{H}_{f}^{-\alpha}\left(\frac{k}{\tilde{a}_{f}\tilde{H}_{f}}\right)^{-2\alpha+1}}{2^{-\alpha+\frac{1}{2}}\Gamma(\frac{3}{2}-\alpha)}
            \\
            d_{2}=\left\{\frac{2^{\frac{1}{2}-\alpha}\left(\frac{k}{\tilde{H}_{f}}\right)^{\alpha}}{\Gamma(\frac{1}{2}+\alpha)}c_{1}\left[\frac{k\left(\frac{k}{\tilde{a}_{f}\tilde{H}_{f}}\right)}{2\alpha+1}+\frac{k\left(\frac{k}{\tilde{a}_{f}\tilde{H}_{f}}\right)^3}{2(2\alpha+1)(2\alpha+3)}\right]\right.
            \\
            \left.\qquad+c_{2}\frac{2^{\alpha-\frac{1}{2}}(-2\alpha+1)(-k)\left(\frac{k}{\tilde{H}_{f}}\right)^{\alpha}}{\Gamma(\frac{3}{2}-\alpha)}\left(\frac{k}{\tilde{a}_{f}\tilde{H}_{f}}\right)^{-2\alpha}\right\}K_{2}^2(\tilde{a}_{f})
        \end{cases}
\end{equation}

And we obtain the post-inflationary magnetic energy density by substituting $\mathcal{A}_{2}(k,\tilde{\eta})$ into Eq.(\ref{e30}) at reheating as,
\begin{eqnarray}
    \frac{d\rho_{B}}{d\ln k}|_{R}\approx\frac{2^{-2\alpha+1}}{32\pi \cos^2(\pi\alpha)}\frac{1}{\tilde{a}_{r}^4\Gamma^2(\frac{1}{2}+\alpha)}\left[\frac{k^{2\alpha+8}\tilde{H}_{f}^{-2\alpha-4}}{(\frac{1}{2}+\alpha)^2\tilde{a}^6_{f}(2\beta+\frac{1}{2})^2}\right]\left(\frac{\tilde{a}}{\tilde{a}_{f}}\right)^{4\beta+1}
\end{eqnarray}

If we assume $\alpha=-2$ which corresponds to a scale invariant magnetic power spectrum during inflation, we obtain the following result,
\begin{eqnarray}
    \frac{d\rho_{B}}{d\ln k}\propto k^4,
\end{eqnarray}
which implies a blue spectrum for magnetic field. We also find that the power spectrum, whose index excluded the parameter $\beta$, is not sensitive to the post-inflationary history\cite{r21}.

\subsection{Magnetic energy density in the Jordan frame after inflation}

To derive the power spectrum of the magnetic field after the end of inflation, it is also necessary to determine the specific form of the coupling function during the reheating stage. The coupling function $I_{2}$ should similarly be obtained through a conformal transformation of $K_{2}$,
and according to the form of $K_{1}$, we can find briskly that $I_{1}$ corresponds to $I(N)$ in Eq.(\ref{e45}).

The expression for $I_{2}$ can be written as,
\begin{equation}
    \begin{split}
        K_{2}&=\left(\frac{\tilde{a}_{f}}{\tilde{a}_{i}}\right)^{-\alpha}\left(\frac{\tilde{a}}{\tilde{a}_{f}}\right)^{-\beta}\\&=\left(\frac{\sqrt{\phi_{1f}}a_{f}}{\sqrt{\phi_{1i}}a_{i}}\right)^{-\alpha}\left(\frac{\sqrt{\phi_{2}}a}{\sqrt{\phi_{1f}}a_{f}}\right)^{-\beta}\\&=I_{1}(N_{f})\left(\frac{\sqrt{\phi_{2}}a}{\sqrt{\phi_{1f}}a_{f}}\right)^{-\beta}=I_{2}
    \end{split}
\end{equation}
where $\phi_{1}$ and $\phi_{2}$ denote the scalar fields during the inflation and reheating periods, respectively, and $N_{f}$ is the number of e-folding at the end of inflation.

Similarly, we demand that $\phi$ and $a$ satisfy continuity at $\eta=\eta_{f}$, in which case $\tilde{a}$ is still continuity at the same time. Thus, we obtained the expression of $a$,
\begin{eqnarray}
    \label{e69}
    a(\eta)=\frac{H_{f}a_{f}^3}{4}\left(\eta+\frac{3}{a_{f}H_{f}}\right)^2
\end{eqnarray}
and as the scalar field amplitude decay causes the root mean square of $\phi$ to decrease with $a^{-3/2}$, converging to a constant on long average, $\sqrt{\phi}$ can be express as\cite{r30,r27},
\begin{eqnarray}
    \sqrt{\phi_{2}}\approx\sqrt{\phi_{1f}}
\end{eqnarray}

The solution of Eq.(\ref{e36}) can also expressed by $a$,
\begin{equation}
    \begin{split}
        \label{e71}
        \bar{A}_{2}(k,\eta)&=D_{1}+D_{2}\int_{\eta_{f}}^{\eta}\frac{1}{I_{2}^2}d\eta\\&=D_1+D_2\int_{a_f}^a\frac{1}{a_{f}^2H_{f}\sqrt{a/a_f}I_{2}^2}da
    \end{split}
\end{equation}
and the expression for $\bar{A}_{1}(k,\eta)$ can be obtained by expanding the Bessel functions in the super-horizon limit as,
\begin{equation}
    \begin{split}
        \bar{A}_{1}(k,\eta)=&C_{1}\frac{\left(\frac{k}{aH_{f}}\right)^{\nu}}{\left(\sqrt{\frac{\phi_{1f}}{\phi_{1i}}}\right)^{\alpha}a_{f}^{\alpha}2^{\nu}\Gamma(1+\nu)}\left[\left(\frac{k}{aH_{f}}\right)^{\frac{1}{2}}-\frac{\left(\frac{k}{aH_{f}}\right)^{\frac{5}{2}}}{4(1+\nu)}+\frac{\left(\frac{k}{aH_{f}}\right)^{\frac{9}{2}}}{32(1+\nu)(2+\nu)}\right]\\&+C_{2}\frac{\left(\frac{k}{aH_{f}}\right)^{-\nu}}{\left(\sqrt{\frac{\phi_{1f}}{\phi_{1i}}}\right)^{\alpha}a_{f}^{\alpha}2^{-\nu}\Gamma(1-\nu)}
    \end{split}
\end{equation}

Then, we also demand that $\bar{A}_{1}(k,\eta)$, $\bar{A}_{2}(k,\eta)$ and their first derivatives are equal at the end of inflation. The coefficients $D_{1}$ and $D_{2}$ can be obtained as,
\begin{equation}
    \begin{cases}
        D_1=C_1\frac{\left(\frac{k}{a_{f}H_{f}}\right)^{\nu}}{\left(\sqrt{\frac{\phi_{1f}}{\phi_{1i}}}\right)^{\alpha}a_{f}^{\alpha}2^{\nu}\Gamma(1+\nu)}\left[\left(\frac{k}{a_{f}H_{f}}\right)^{\frac{1}{2}}-\frac{\left(\frac{k}{a_{f}H_{f}}\right)^{\frac{5}{2}}}{4(1+\nu)}+\frac{\left(\frac{k}{a_{f}H_{f}}\right)^{\frac{9}{2}}}{32(1+\nu)(2+\nu)}\right]
        \\
        \qquad+C_{2}\frac{\left(\frac{k}{a_{f}H_{f}}\right)^{-\nu}}{\left(\sqrt{\frac{\phi_{1f}}{\phi_{1i}}}\right)^{\alpha}a_{f}^{\alpha}2^{-\nu}\Gamma(1-\nu)}
        \\
        D_2=\left\{\frac{C_1}{\left(\sqrt{\frac{\phi_{1f}}{\phi_{1i}}}\right)^{\alpha}a_{f}^{\alpha}2^{\nu}\Gamma(1+\nu)}\left[(-k)(\nu+1/2)(\frac{k}{a_{f}H_{f}})^{\nu-1/2}-
            \frac{-k(\nu+5/2)(\frac{k}{a_{f}H_{f}})^{\nu+3/2}}{4(1+\nu)}\right.\right.
            \\
            \left.\left.\qquad+\frac{-k(\nu+9/2)(\frac{k}{a_{f}H_{f}})^{\nu+7/2}}{32(1+\nu)(2+\nu)}\right]+\frac{C_2}{\left(\sqrt{\frac{\phi_{1f}}{\phi_{1i}}}\right)^{\alpha}a_{f}^{\alpha}2^{-\nu}\Gamma(1-\nu)}\left[k\nu\left(\frac{k}{a_{f}H_{f}}\right)^{-\nu-1}\right]\right\}I_{2}^2(a_{f})
    \end{cases}
\end{equation}

Substituting $D_1$ and $D_2$ into Eq.(\ref{e71}), we get energy density of magnetic field by using Eq.(\ref{e37}),
\begin{equation}
    \begin{split}
        \frac{d\rho_{B}}{d\ln k}|_{R}\approx&\frac{2^{2\nu}}{8\pi \cos^2[\pi(\nu+1/2)]}\frac{1}{a_{r}^4\Gamma^2(1-\nu)\left(\sqrt{\frac{\phi_{1f}}{\phi_{1i}}}\right)^{2\alpha}}\\&\times\left[\frac{k^{-2\nu+4}H_{f}^{2\nu}}{\nu^{-2}a_{f}^{\alpha-\nu+4}(2\beta+\frac{1}{2})^2}\right]\left(\frac{a}{a_{f}}\right)^{4\beta+1}
    \end{split}
\end{equation}
where $a_{r}$ is the scale factor in the Jordan frame at the end of reheating.

Assuming $\nu=\alpha=-2$ (the scale-invariant spectrum in the Einstein frame), we obtain $d\rho_{B}(k,\eta)/d~\ln
k\propto k^8$. This implies that the energy spectrum of the magnetic field follows a blue spectrum. Moreover, since the exponent of $k$ does not include the parameter $\beta$, it indicates that the inflationary magnetogenesis is not sensitive to the post-inflationary history in the Jordan frame.

\section{SUMMARY\label{s5}}
In summary, The issue of physical equivalence between the Einstein frame and the Jordan frame remains an unresolved question. In the context of primordial magnetic field cosmology, this paper calculates the energy density and scale-invariant spectrum of magnetic fields in both frames during the inflationary epoch and the reheating phase. A simple $f(R)$ gravity theory, specifically the $R^{2}$ gravity theory, is selected for this study. Since the two frames are connected via a conformal transformation, we assume a power-law coupling in the Einstein frame and derive the corresponding coupling function in the Jordan frame through the conformal transformation (which is no longer a power-law in its conformal form). Through calculations, we find that during inflation, the energy density and scale-invariant spectrum of the magnetic fields differ between the two frames. If physical equivalence between the frames is required, the coupling function must be a constant. During the reheating phase, we find that for ( $\nu=\alpha= -2$ ), the magnetic fields in both frames exhibit a blue spectrum (with higher energy density in the Jordan frame), and both are independent of the post-inflationary history.

We hope that our results provide some insights into the physical equivalence between the two frames. Similar analyses could also be conducted for other $f(R)$ theories, which opens up possibilities for future work.

\section*{Acknowledgments}
This work was supported by the Fundamental Research Funds for the Central Universities of Ministry of Education of China under Grant No.3132018242, the Natural Science Foundation of Liaoning Province of China under Grant No.20170520161 and the National Natural Science Foundation of China under Grant No.11447198 (Fund of theoretical physics).

%\begin{thebibliography}{000} %for 3 digits
%\begin{thebibliography}{00}  %for 2 digits
\bibliographystyle{ws-ijmpd}
\bibliography{Wh-ijmpd}

\end{document}